\documentclass[aps,prd,twocolumn,nofootinbib,showpacs,floatfix,superscriptaddress]{revtex4}
\usepackage{amsmath, amsthm}
\usepackage[dvips]{graphicx}

\newcommand{\lsim}{\lower0.6ex\vbox{\hbox{$ \buildrel{\textstyle <}\over{\sim}\ $}}}
\newcommand{\gsim}{\lower0.6ex\vbox{\hbox{$ \buildrel{\textstyle >}\over{\sim}\ $}}}
\newcommand{\beq}{\begin{equation}}
\newcommand{\eeq}{\end{equation}}

\begin{document}


\title{
Lyman-$\alpha$ Forest Constraints on Decaying Dark Matter
}


\author{Mei-Yu Wang}
\email[E-mail : ]{meiywang@indiana.edu} 
\affiliation{Department of Physics, Indiana University at Bloomington, Bloomington, IN 47405-7105}
\affiliation{Department of Physics and Astronomy \& Pittsburgh Particle Physics, Astrophysics, and Cosmology Center (PITT PACC), University of Pittsburgh, Pittsburgh, PA 15260}
\author{Rupert A.C. Croft}
\affiliation{Department of Physics, Carnegie Mellon University, Pittsburgh, PA 15213}
\author{Annika H. G. Peter}
\affiliation{CCAPP and Departments of Physics and Astronomy, The Ohio State University, Columbus, OH 43210}
\affiliation{Center for Cosmology, Department of Physics and Astronomy, University of California, Irvine, CA 92697-4575}
\author{Andrew R. Zentner}
\affiliation{Department of Physics and Astronomy \& Pittsburgh Particle Physics, Astrophysics, and Cosmology Center (PITT PACC), University of Pittsburgh, Pittsburgh, PA 15260}
\author{Chris W. Purcell}
\affiliation{Department of Physics, West Virginia University, Morgantown, WV 26506-6315}
\affiliation{Department of Physics and Astronomy \& Pittsburgh Particle Physics, Astrophysics, and Cosmology Center (PITT PACC), University of Pittsburgh, Pittsburgh, PA 15260}


\begin{abstract}
We present an analysis of high-resolution N-body simulations of decaying dark matter cosmologies focusing 
on the statistical properties of the transmitted Lyman-$\alpha$ (Ly$\alpha$) forest flux in the high-redshift intergalactic medium (IGM). 
In this type of model a dark matter particle decays into a slightly less massive stable dark matter daughter particle and a comparably light particle. 
The small mass splitting provides a non-relativistic kick velocity $V_k = c\, \Delta M/M$ to the daughter particle resulting in free-streaming 
and subsequent damping of small-scale density fluctuations. Current  Ly$\alpha$ forest power spectrum measurements probe 
comoving scales up to  $\sim 2-3\, h^{-1}$~Mpc at redshifts $z \sim 2-4$, providing one of the most robust ways to probe cosmological 
density fluctuations on relatively small scales. The suppression of structure growth due to the free-streaming of dark matter daughter particles 
also has a significant impact on the neutral hydrogen cloud distribution, which traces the underlying dark matter distribution well at high redshift. 
We exploit Ly$\alpha$ forest power spectrum measurements to constrain the amount of free-streaming of dark matter in such models and 
thereby place limits on decaying dark matter based only on the dynamics of cosmological perturbations without any assumptions about the 
interactions of the decay products. We find that SDSS 1D Ly$\alpha$ forest power spectrum data place a lifetime-dependent 
upper limit $V_k \lesssim 30-70$~km/s for decay lifetimes $\lesssim 10~\mathrm{Gyr}$. This is the most stringent model-independent 
bound on invisible dark matter decays with small mass splittings. For large mass splittings (large $V_k$), Ly$\alpha$ forest data restrict 
the dark matter lifetime to $\Gamma^{-1} \gtrsim 40~\mathrm{Gyr}$. Forthcoming BOSS data should be able to provide more stringent 
constraints on exotic dark matter, mainly because the larger BOSS quasar spectrum sample will significantly reduce statistical errors. 
\end{abstract}

\date{\today}
\maketitle

\section{Introduction}
\label{section:introduction}
The formation of structure in the universe is driven by the mysterious dark matter (DM) component whose nature remains unknown. Over the last few decades, the hierarchical cold dark matter model (CDM) has become the standard description for the formation of cosmic structures. The most popular candidate for 
CDM is the class of weakly-interacting massive particles (WIMPs), among which is the lightest neutralino in supersymmetric extensions to the standard model of particle physics \cite{jungman_etal96b,griest_kamionkowski00,bertone_etal05}. In this paper, we describe generic constraints placed by observations of the large-scale structure of the universe on alternative scenarios in which the dark matter particle decays invisibly with a long lifetime ($\gtrsim \mathrm{Gyr}$). The constraints that we derive are competitive with constraints derived from galactic substructure and represent the most stringent, model-independent constraints on unstable dark matter.

The CDM model is consistent with the cosmic microwave background (CMB) anisotropy  spectrum measured by the Wilkinson Microwave Anisotropy Probe (WMAP) \cite{WMAP9} and PLANCK \cite{Planck_13} and observations of the large-scale ($k \lesssim 0.1$~h/Mpc) galaxy clustering spectrum measured by the Sloan Digital Sky Survey (SDSS) \cite{Tegmark_etal06}. Moreover, CDM has significant predictive power because the only parameter describing the dark matter in this theory, 
$\Omega_m$, is now very well constrained by observational data (e.g., \cite{Planck_13}). However, there are a number of observations on sub-galactic scales that may challenge the CDM paradigm. Among these possible tensions are: (1) an excess in the predicted number of galactic satellites \cite{klypin_etal99b,moore_etal99}; (2) inferred central densities of observed low-surface brightness (LSB) galaxies that appear to be smaller than predicted by CDM \cite{deBlok_etal02,zentner_bullock02,Simon_etal05,KuziodeNaray_etal08,Oh_etal11}; and (3) the kinematics of the 
Milky Way's massive dwarf satellites indicate that they are underdense compared to predictions for the largest satellite galaxies \cite{Boylan-Kolchin_etal11}.  There are a number of plausible explanations for these discrepancies within the CDM paradigm. For example, the over-prediction of the number of galactic satellites may be explained by baryonic feedback (Ref.~\cite{brooks_etal13} is a recent example among numerous papers exploring this possibility), the apparent low central densities of LSB galaxies may reflect both baryonic feedback (e.g., \cite{Governato_etal12,Teyssier_etal13}) and references therein) as well as halo triaxiality and non-circular motions in these galaxies (e.g., Ref.~\cite{hayashi_etal07}, but see the counter example in Ref.~\cite{naray_etal11}), and the densities of the largest satellite galaxies can easily be explained by halo-to-halo variation \cite{purcell_zentner12}, a Milky Way halo mass that is slightly smaller than assumed in many studies \cite{wang_etal12}, or both.

Nonetheless, these studies have stimulated significant exploration of alternative models of dark matter and it is interesting to explore foils to the CDM paradigm. Some thermal relic DM candidates, such as sterile neutrinos and gravitinos, have a mass of order one keV with non-negligible thermal velocities that suppress structure smaller than their free-streaming scales. Such warm dark matter (WDM) particles can also be produced via {\it resonant oscillations} which result in models with a  mixture of cold+warm DM (CWDM) \cite{boyarsky_etal08}. Alternatively, self-interacting dark matter (SIDM) models, in which dark matter particles interact with each other through large cross sections, have renewed attention recently and such scenarios may be realized in hidden-sector extensions to the Standard Model  \cite{Spergel_etal00,Arkani-Hamed_etal09,Feng_etal09,Loeb_etal11}. As another alternative to CDM, unstable dark matter has been considered in a number of recent studies as well (e.g., \cite{peter_10,peter_benson10,peter_etal10,wang_zentner12}). A recent review of the current status of solutions to small-scale problems can be found in \cite{Weinberg_etal13}.

The most restrictive constraints on thermal WDM models come from Ly$\alpha$ forest measurements \cite{Viel_etal05,Abazajian_06,Seljak_etal06,boyarsky_etal08,Viel_etal13}. Recent studies have shown that, within the allowed parameter range, WDM may not suffice  to solve the cusp-core problem and may not reduce small-scale power sufficiently to reduce the satellite population in Milky Way halos \cite{Polisensky_etal11, Maccio_etal12}. Resonantly produced WDM with a mixture of cold and warm components may still be a viable explanation for the small-scale issues \cite{Maccio_etal12b}. For SIDM, there remains a very narrow window of viable model parameters (consistent with
observations) that may give rise to cored density profiles and mitigate the problem of the densities of the largest satellite galaxies of the 
Milky Way. Interestingly, SIDM does not significantly reduce the number of satellites in Milky Way, in contrast to WDM models, and so there is a 
distinct possibility that these models can be distinguished from each other through observations of galaxies in the Local Group\cite{Vogelsberger_etal12,Rocha_etal12,Peter_etal13}.

In models of unstable DM, a DM particle of mass $M$ decays into a less massive, stable daughter particle of mass
$m = (1-f)M$ and a significantly lighter, relativistic particle, with a lifetime on the order of the age of the
universe. The stable daughter particle will recoil to conserve momentum in the decay and the magnitude of the recoil speed is proportional to the phenomenological mass-splitting fraction $f$. As a result of decays, the DM is effectively a mixture of cold and warm components (the cold parents and the warm daughters), a situation that is broadly similar to the CWDM models produced through resonant oscillations. The linear matter power spectrum in this class of decaying dark matter (DDM) is characterized by a step and a plateau on small scales \cite{wang_zentner12}. In this regard, the low-redshift phenomenology in the DDM model is broadly similar to the power spectra of WDM and CWDM \cite{boyarsky_etal08}; however, the late-time evolution of the power spectrum is more significant in DDM models because the dark matter decays occur in the contemporary universe so that the warm component of the dark matter is produced only at low redshift \cite{wang_zentner12}. The late-time evolution of structure in DDM models renders them distinguishable from WDM and CWDM scenarios in principle. All of these features show that DDM might provide another plausible mechanism for mitigating some of the small-scale issues faced by CDM. Therefore, it is worth investigating current limits on DDM models.

Following the methods that have been applied to study WDM and CWDM, in this study we consider constraints 
on DDM models from current large-scale structure measurements such as the Ly$\alpha$ forest. The Lyman-absorption features produced by 
intergalactic neutral hydrogen clouds in the spectra of distant quasars, which are collectively called the Lyman-$\alpha$ forest, 
have been shown to trace the dark matter distribution closely in simulated data \cite{croft1997}. The current Ly$\alpha$ forest data can probe 
one-dimensional (1D) fluctuations as small as $\sim$~Mpc at redshifts z$\sim$2-4 \citep{kim_etal04,Croft_etal02,McDonald_etal06}. Recently the Baryon Oscillation and 
Spectroscopic Survey (BOSS) project released a measurement of 1D Ly$\alpha$ forest power spectrum data \cite{Palanque-Delabrouille_etal13} which greatly reduces the statistical error over the previous SDSS data set \cite{McDonald_etal06} by utilizing a larger quasar spectrum sample in their analysis. In our study, we derive robust, model-independent constraints on DDM based on current SDSS data \cite{McDonald_etal06} and the derived matter power spectrum from high resolution Ly$\alpha$ forest measurements \cite{kim_etal04,Croft_etal02,VHS_04}, and discuss possible improvements from BOSS data. In our follow-up work, we use zoom-in numerical 
simulations to explore the ability of DDM models to solve the small-scale problems in galactic halos (Wang et al., in preparation). We find 
that DDM models severely constrained by extant data, yet DDM may provide a plausible solution to some of the small-scale problems of CDM 
within the allowed parameter range.

The outline of this paper is as follows. In \S~\ref{section:models}, we briefly review the perturbative evolution in DDM models. In \S~\ref{section:methods} we introduce the two Ly$\alpha$ forest data sets used in this analysis: 
the Viel, Haenelt $\&$ Springel (VHS) data compiled in \cite{VHS_04}, and the SDSS  data presented by \citep{McDonald_etal06}. We 
describe our simulations, which are aimed at the range of scales probed by the SDSS 1D flux power spectrum, in \S~\ref{section:simulations}. 
In \S~\ref{section:PF} we describe the framework for our flux power spectrum reconstruction. The results, including simulated matter and flux power spectra 
as well as the limits on DDM models, are presented in \S~\ref{section:results}. We conclude in \S~\ref{section:conclusion}.

\section{Decaying Dark Matter Models}
\label{section:models}

In this section, we describe briefly the parameters of DDM decaying into invisible particles and review the 
phenomenology of structure formation in such models. We refer readers to \cite{wang_zentner12} 
for detailed discussions and derivations related to this class of unstable DM. 

We consider DM decays into another species of stable 
DM (SDM) with a small mass splitting, DDM~$\rightarrow$~SDM~$+$~L, 
where L denotes a light daughter particle the mass of which is significantly 
below the mass of the SDM, so that the light daughter particle is relativistic after the decay event. 
SDM is the stable DM with mass $m$, and DDM is the decaying DM with mass $M$. 
The mass loss fraction $f=(M-m)/M$, of DDM is directly related to the recoil kick velocity 
of the SDM particle by $f \simeq V_k/c$ from energy-momentum conservation 
(assuming $V_k/c \ll 1$ which is true over most of the viable yet interesting 
parameter space). 

There are two parameters in this class of DDM models: decay rate $\Gamma$ (or decay lifetime $\Gamma^{-1}$); and kick velocity $V_k$ (or mass splitting fraction $f$). As we will describe later, the relevant decay lifetimes in this work are generally large, ranging from $\sim 0.1$~Gyr to a many Hubble times. The advantage of the late time decay model is that it can provide a possible solution to the small scale problems observed at present without altering the successes of CDM models on large-scales and at early times. In the limit that the DDM lifetime is very short compared to the Hubble time, the behavior of the DDM will be similar to WDM with the mass splitting setting the velocity scale and, therefore, playing a role analogous to the WDM mass. With large decay lifetime, the difference between DDM and WDM is the evolution of the free-streaming scale as a function of time. While for WDM (also applicable to standard model neutrinos) the free-streaming scale gradually shrinks after the WDM particles decouple, the late-decaying DM model keeps generating particles with excess kick velocities, causing the free-streaming scale to expand until late times (see Fig. 3 in \cite{wang_zentner12} for a detailed comparison). 

The conversion of DM into relativistic energy is suppressed by a factor which
is the mass difference between the DDM particles and SDM particles. However, later we will see that in our study the relevant kick velocity range for SDM particles, which are generated from the mass difference, is around a few hundred km/s or less. Therefore, the fraction of mass-energy that is converted into relativistic energy is suppressed by a factor of order $\sim 10^{-3}$. This suppression renders the effect of decays on the universe's expansion history sufficiently small as to be unobservable with contemporary data, but the excess velocity imparted to the SDM particles can result in significant imprints on matter density perturbations. These imprints can be examined using observational data that probe structure growth, such as future weak lensing surveys \cite{wang_zentner12} and, in this work, the 
Ly$\alpha$ forest.

\section{Lyman-$\alpha$ Forest methods}
\label{section:methods}
The Ly$\alpha$ forest from redshifts $z \sim 2-4$
is a dense set of absorption features seen in quasi-stellar object (QSO) spectra. 
These features are caused by the residual neutral hydrogen present in a photoionized IGM.
The gaseous structures responsible for typical Ly$\alpha$ forest lines are large ($\ge$ 100 kpc), low density ($\delta\rho/\rho \le $ 10), and 
fairly cool (T $\sim 10^4$ K), so pressure forces are sub-dominant, and the gas density closely traces the total matter density on large scales. 
The Fourier-transformed Ly$\alpha$ forest flux power spectrum $P_F(k,z)$ thus provides a way to estimate the matter fluctuation on scales up to 
$k \sim$~a few~$h\mathrm{Mpc}^{-1}$ at high redshift.

The flux power spectrum $P_F(k)$ and linear matter matter power spectrum $P_m(k)$ are complicated functions of cosmology as well as astrophysical parameters that are related to properties of intergalactic gas. In order to alleviate degeneracies of DDM model parameters with cosmological and astrophysical parameters, we include current CMB and large-scale galaxy power spectrum data in the constraints we derive. The parameter limit extraction from combined cosmological data sets can be conveniently performed with Markov Chain Monte Carlo  (MCMC) techniques using the public code {\tt CosmoMC} \cite{Lewis_etal02}. In the following, we present the two Ly$\alpha$ data sets that we have adopted in our analysis and the approaches we take to utilize them.

\subsection{VHS Data}
\label{subsection:VHS}

The VHS data \citep*{VHS_04} contain two high signal-to-noise sets of spectra: one has 27 QSO spectra from LUQAS (Large Sample of UVES QSO Absorption Spectra) by \cite{kim_etal04} with mean absorption redshift $z$ $\sim$ 2.25, and the other one from \cite{Croft_etal02} consists of 30 Keck HIRES spectra and 23 Keck LRIS spectra with $z$ $\sim$ 2.72 between 2.3 $<$ z $<$ 3.2. 

These data are analyzed in \citep*{VHS_04}  using a large suite of hydrodynamic simulations to map the 1D flux power spectrum $P_F(k,z)$ into the matter power spectrum $P_m(k,z)$ on scales 0.003 s/km $\lsim$ k $\lsim$ 0.03 s/km, which roughly corresponds to scales 0.3 h/Mpc $\lsim$ k $\lsim$3 h/Mpc. The bias functions $b^2(k,z)=P_F(k,z)/P_m(k,z)$ as seen in \cite{Viel_etal05} differ very little between the WDM and CDM scenarios over the relevant range of wavenumbers. Due to the similarity of WDM and DDM behavior in suppressing structure growth, we take the same approach as these authors and derive limits from this dataset using the derived matter power spectrum from \citep*{VHS_04}. In principle, more rigorous limits require a large suite of hydrodynamic simulations to explore how the bias function changes with DDM model parameters as well as cosmological and astrophysical parameters, but the Ref.~\cite{Viel_etal05} suggests that this is unnecessary at this level of precision. 

We vary five standard cosmological parameters, namely the contemporary dark matter density parameter $\Omega_m h^2$ (in principle $\Omega_m$ varies in a non-trivial manner in DDM scenarios, but in cases of interest this effect is negligibly small), the baryon density parameter $\Omega_b h^2$, the power spectrum normalization probed by the CMB anisotropy $A_s$, the power-law index of the scalar perturbation spectrum $n_s$, and the optical depth due to reionization $\tau_{re}$. We also vary the two parameters of DDM models which are the lifetime $\Gamma^{-1}$ of the DDM particles and kick velocity $V_{k}$ given to the decay products (or, equivalently, the parent-daughter mass splitting). We utilize the VHS Ly$\alpha$ forest module in {\tt CosmoMC} \cite{Lewis_etal02} combining with modified {\tt CAMB} to derive limits on decay parameters. Additional data included in our analysis in order to mitigate parameter degeneracies are 
the SDSS galaxy 3D power spectrum \citep{Tegmark_etal06} and CMB constraints from seven year WMAP experiment \citep{Komatsu_etal11}. 

\subsection{SDSS Data}
\label{subsection:SDSS}

The SDSS collaboration \citep{McDonald_etal06} has analyzed 3035 quasar spectra with relatively low resolution and low signal-to-noise. They span a wide redshift range from $z \sim 2.2$ to $z \sim 4.2$. The resolution is too low to provide measurements on small scales (k $\ge$ 0.02 s/km), but the large sample number significantly reduces the statistical error on large scales and compensates for the low signal-to-noise. Consequently it is not possible to neglect the effects of cosmological parameters on the bias function in an analysis of this data. In what follows, we work with the 1D flux power spectra directly. We take advantage of the tight correlations between temperature and density expected in the IGM and run a suite of $N$-body simulations to approximate these effects. A more robust result would require running full hydrodynamic simulations including galaxy formation and a treatment of radiative transfer, something that is beyond the scope of this project. In this work, we derive constraints based upon well-supported approximations as a proof of concept while we continue to hone in on the region of DDM parameter space that is simultaneously viable and of astrophysical/cosmological interest. We describe our simulations in \S~\ref{section:simulations} and our methods for estimating flux power spectra from the simulation data in \S~\ref{section:PF}

Following the approach of \cite{Viel_etal06}, we have approximated the flux power spectrum by a first-order Taylor expansion for the cosmological/astronomical parameter vector $\textbf{p}$ around the fiducial model $\textbf{p}^0$:
\beq
\label{Taylor}
P_{F}(k,z;\mathbf{p})=P_{F}(k,z;\mathbf{p^0})+\sum^N_i {\partial P_{F} (k,z;p_i)\over \partial p_i}\Big\arrowvert_{\mathbf{p}=\mathbf{p^0}} (p_i-p_i^0),
\eeq
where $p_i$ are the N components of the vector $\textbf{p}$.
We then perform a MCMC analysis in this parameter space to take into account the uncertainties associated with them. We have the same set of cosmological parameters as in the VHS analysis: $\Omega_m h^2$, $\Omega_b h^2$, $A_s$, $n_s$, $\tau_{re}$,  $\Gamma^{-1}$, and $V_{k}$. For astrophysical nuisance parameters we consider eight parameters that deal with uncertainties related to Ly$\alpha$ physics, as suggested by \cite{Viel_etal06}. We also have 9 additional parameters that model a number of corrections to the data following the suggestions in \citep{McDonald_etal06}. We explain in detail how we apply the nuisance parameters to the simulated flux power spectrum in \S~\ref{section:PF}.

\section{Simulations}
\label{section:simulations}
In order to predict $P_F(k,z)$ for a given cosmological model, we have performed $N$-body simulations of cosmological structure growth and applied the {\it fluctuating Gunn-Peterson approximation} \cite{Croft_etal98,Gnedin_etal98} to create Ly$\alpha$ forest spectra assuming that gas and DM have the same spatial distribution at high redshift. Our goal is to estimate the first Ly$\alpha$ constraints on the DDM models and we have adopted this approach in order to conserve computational resources in this proof-of-concept study. This approximation has been shown \citep{Croft_etal98, Gnedin_etal98, Meiksin_etal01} to reproduce results comparable to the full hydrodynamical simulations with errors $\lsim$10$\%$. In the future, to utilize the more precise BOSS data \cite{Palanque-Delabrouille_etal13}, it will be necessary to perform full hydrodynamic simulations that include a number of additional baryonic effects. The recently-developed LyMAS method \cite{Peirani} for predicting clustering statistics using a combination of hydrodynamic and dark matter simulations may also prove to be a suitable alternative.

We used a version of the parallel N-body code GADGET-2 \cite{Springel_etal05} modified by \citep{peter_etal10} to simulate the effects of 
dark matter decay on the Ly$\alpha$ forest. This code version includes a Monte-Carlo simulation at each time step $\Delta t$ to determine whether a particle should decay with decay probability $P=\Gamma\Delta t$ . If a particle is designated for decay, it will receive a kick speed $V_{k}$ in a random direction, and it is flagged to make sure it will not decay again. Each simulation is carried out in a cubic periodic box of size of 60$h^{-1}$~Mpc on a side, using $400^3$ DM particles. The gravitational softening scale is set to 1~$h^{-1}$ kpc in comoving units and the mass per particle is 2.56$\times 10^8$ $h^{-1} M_{\odot}$. All the decay simulations have the same initial Fourier phases as the fiducial run, starting at z=99. Snapshots are output at 11 regularly spaced redshift intervals between z=4.2 and 2.2. The initial conditions are generated using N-GenIC by displacing particles from a Cartesian grid according to the Zel'dovich approximation to obtain distributions that agree with the density fluctuation power spectrum from \citep{Eisenstein_hu_98}.

To map simulation results into parameter constraints, we adopt the first-order Taylor expansion method (see Eq. \ref{Taylor}) from \cite{Viel_etal06}. Although this method will become inaccurate when the points are far from the fiducial model and it assumes that the likelihood distributions are well-described by multivariate Gaussian functions, it has been found \citep{Viel_etal06, boyarsky_etal08} to be a good approximation for the standard cosmological/astronomical parameters. We emulate the approach of \cite{boyarsky_etal08} (applied to CWDM models) to treat our decay lifetime and velocity kick parameters and we run a grid of simulations to sample the decay model parameter space, which is highly non-Gaussian. We perform a set of 16 DDM simulations, with $\Gamma^{-1}$ = 30, 10, 1, 0.1 (Gyr) and $V_{k}$ = 70, 100, 200, 500 (km/s) to cover the decay parameter space. Then we apply extrapolation among the simulation grids to derive the decay model predictions. The cosmological reference model corresponds to a "fiducial" $\Lambda$CDM universe with parameters at z=0, $\Omega_m$=0.273, $\Omega_{\Lambda}$=0.727, $\Omega_b$=0.044, $n_s$=0.967, $H_0$=70.4 km$s^{-1}Mpc^{-1}$, and $\sigma_8$=0.811. It is consistent with the results of the WMAP 7-year data analysis \citep{Komatsu_etal11}. We also run additional simulations that vary $H_0$, $n_s$, $\sigma_8$, $\Omega_m$ to calculate the power spectrum differences induced by changes in these parameters.

\section{The Flux Power Spectrum}
\label{section:PF}

From previous Ly$\alpha$ forest studies using numerical simulations, it has been found that the relation between 
temperature and density is well-approximated by a power law:
\beq
\label{t_rho}
T=T_0(\rho_b/\bar{\rho_b})^{1-\gamma},
\eeq
where $T_0$ is of the order of $10^4$ K.
We can derive the optical depth $\tau$ by applying  the ``fluctuating Gunn-Peterson approximation'' \cite{Croft_etal98,Gnedin_etal98}, giving 
\beq
\label{tau_rho}
\tau \propto \rho^2_bT^{-0.7}=A(\rho_b/\bar{\rho_b})^{\beta}=A(\rho_{DM}/\bar{\rho_{DM}})^{\beta},
\eeq
where
\begin{align}
\label{FGPA}
A=&0.946 \Big({1+z\over4}\Big)^6\Big({\Omega_bh^2 \over 0.0125}\Big)^2\Big({T_0\over10^4~\mathrm{K}}\Big)^{-0.7}\Big({\Gamma_{photo}\over 10^{-12} ~\mathrm{s}^{-1}}\Big)^{-1} \\
& \times\Big({H(z)\over 100~{km/s/Mpc}}\Big)^{-1},
\end{align}
with $\beta \equiv$ 1.3+0.7$\gamma$. Here $\Gamma_{photo}$ is the photoionization rate and $H(z)$ 
is the Hubble expansion rate at redshift $z$. The flux is calculated by
\beq
\label{tau_f}
\mathcal{F}=e^{-\tau},
\eeq
The flux power spectrum is defined by
\begin{align}
\label{pf}
P_F(k) &= \big | \delta_F(k)\big |^2  \\
\delta_F&={\mathcal{F} \over \bar{\mathcal{F}}} -1,
\end{align}
where $\bar{\mathcal{F}}(z)=exp(-\tau_{eff}(z))$.

Each simulation snapshot was processed to generate an averaged flux power spectrum as follows. 
First, 10,000 randomly-placed simulated quasar sightlines were drawn through out a simulation box. 
For each line-of-sight Equation \ref{tau_rho} was used to generate skewers of optical depth with 2500 pixels each. 
The $\tau$ values on the skewers were then convolved with the line-of-sight
$\tau-$weighted velocity field to produce a redshift space optical depth
field. This was then converted to transmitted flux using Equation \ref{tau_f}.

\begin{figure*}[ht]
\includegraphics[height=7.5cm]{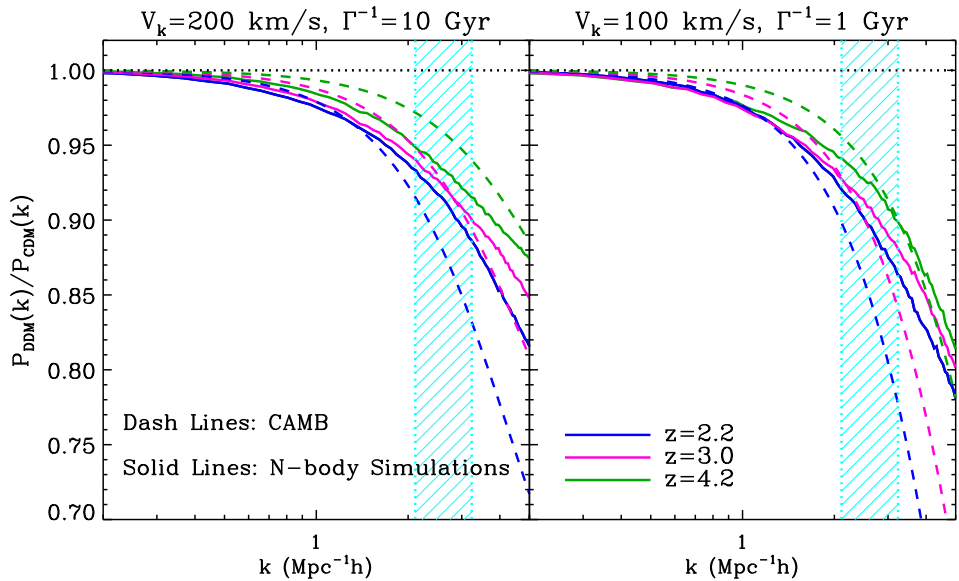}
\caption{ 
Ratio of DDM 3D matter power  spectrum relative to CDM in the relevant parameter range. In the left panel we show a comparison of power spectrum difference generated from a modified {\tt CAMB} code (dashed lines) with that from DDM N-body simulations with decay parameters $V_{k} $=200 km/s and $\Gamma^{-1}$=10 Gyr at redshift z=2.2, 3.0, and 4.2 (bottom to top). The right panel shows the same comparison with $V_{k} $=100 km/s and $\Gamma^{-1}$=1 Gyr. The light blue shaded areas indicate the upper limit on $k$ for SDSS Ly$\alpha$ forest data and VHS data.
}
\label{fig:PK_3d}
\end{figure*}

To marginalize over the uncertainties in $\tau_{eff}$, $T_0$, and $\gamma$, we use nine astrophysical parameters, following the methods of \cite{Viel_etal06}. 
For $\tau_{eff}$, we have amplitude $(\tau_{eff}^A)$ and slope at z=3$(\tau_{eff}^S)$, so that the evolution of effective optical depth is described as a power-law: $\tau_{eff}(z)=\tau_{eff}^A[(1+z)/4]^{\tau_{eff}^S}$. We treat both $\gamma$ and $T_0$ as broken power-laws at z=3 with one amplitude parameter and two slopes at 
$z < 3$ and $z > 3 $. The three parameters for the temperature are the amplitude at z=3, $T_0^A(z=3)$ and the two slopes $T_0^S(z<3)$ and  $T_0^S(z>3)$. The parameter $\gamma$ is described in the same fashion with $\gamma^A(z=3)$, $\gamma^S(z<3)$, and $\gamma^S(z>3)$. 

We also have 9 additional parameters that model a number of corrections to the data. Following the suggestions in \citep{McDonald_etal06}, we have seven parameters $f_i, i=1-7$ for the noise correction in redshift bins at z=2.2 - 3.4. In each redshift bin we subtract $f_i P_{noise}(k, z)$ from $P_f (k, z)$ and treat $f_i$ as free parameters with the assumption of Gaussian distributions. To allow for the overall resolution correction, we multiply $P_F (k, z)$ by exp($\alpha k^2$), where $\alpha$ is treated as a free parameter with a Gaussian distribution. We also account for the presence of damped Ly$\alpha$ system by fitting $A_{damp}$ following the suggestions in \cite{McDonald_etal05} and \cite{Viel_etal06}.

Using the method of \citep{McDonald_etal06}, we include a model for the contamination of the signal by Si III lines by assuming a linear bias correction of the form $P_{F}'(k)=[(1+a^2)+2a$ cos$(vk)] P_F(k)$, where $a$ = $f_{SiIII}/(1-\bar{\mathcal{F}}(z))$ with $f_{SiIII}$ = 0.011 and $v$ = 2271 km/s. We also include the full data covariance matrix in our MCMC analysis.

\begin{figure*}[ht]
\includegraphics[height=10.0cm]{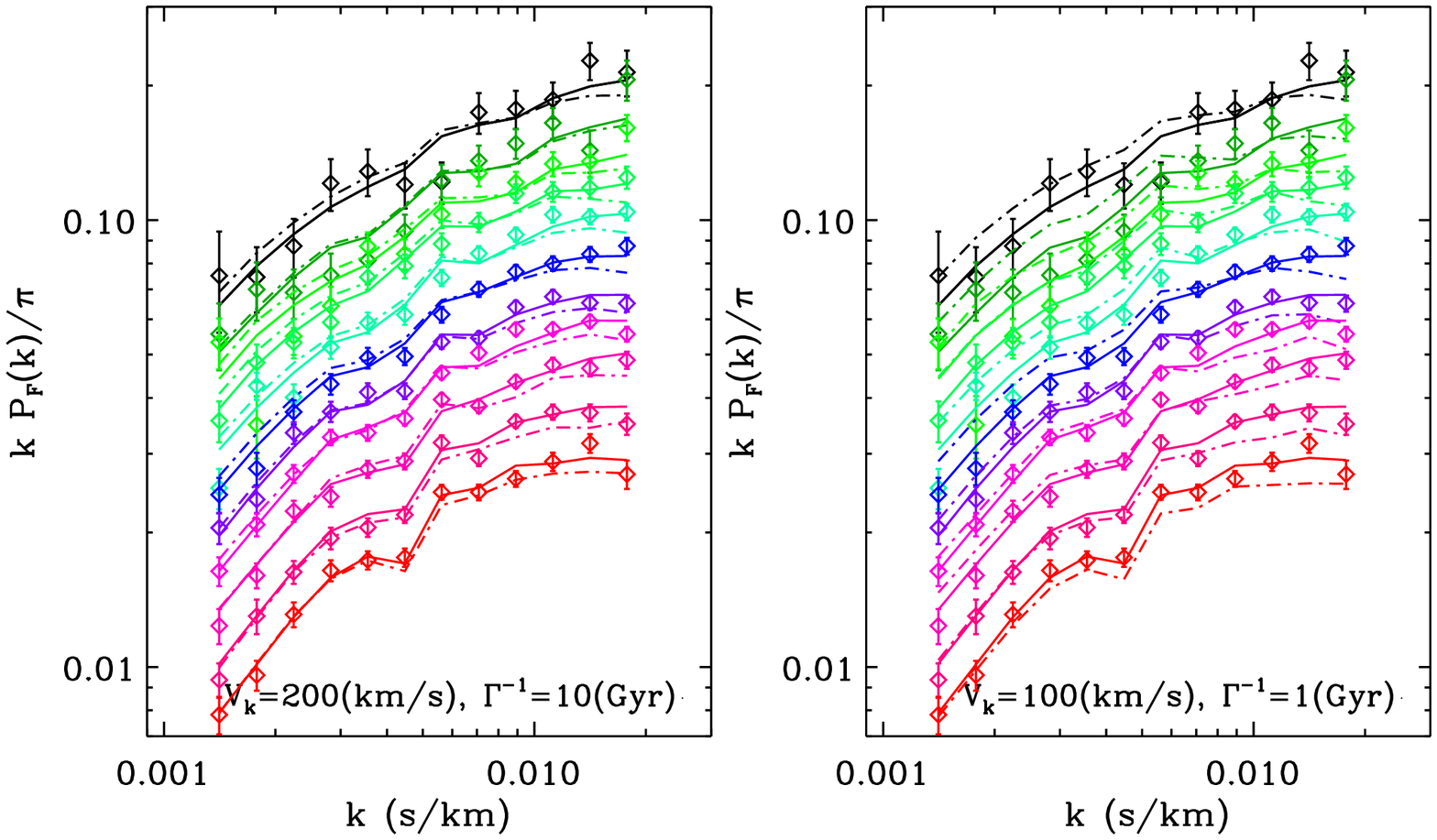}
\caption{ 
Comparison of the SDSS 1D Ly$\alpha$ forest power spectra \citep{McDonald_etal06} as a function of redshift from z=4.2 (top black diamond points with 1 $\sigma$ error bar) to 2.2 (bottom red diamond points) in steps of 0.2 with predictions from our numerical simulations. For each redshift bin the solid lines are from the best-fit CDM model simulations, while the dash-dotted lines are from DDM simulations with the decay parameter value marked in each panel. $V_k$ is the recoil kick velocity, and $\Gamma^{-1}$ is the decay lifetime. 
}
\label{fig:FP}
\end{figure*}

\section{Results}
\label{section:results}

\subsection{The Matter Power Spectrum in Decaying Dark Matter Models}
\label{subsection:linear power spectrum}

There are several effects which DDM has on structure growth at high redshift. 
We briefly mention some of the important features here. A detailed study can be found in \cite{wang_zentner12}. 

First, the decay process converts some of the DM  into relativistic energy and thus changes the evolution of the cosmological energy densities. In principle, this change alters both structure growth and the cosmological distance-redshift relation. However, as we mention in \S~\ref{section:models}, this is an effect that can be neglected in this study. We will see that in \S~\ref{section:constraints} the most relevant decay parameter values constrained by the Ly$\alpha$ forest include $V_{k}$ $\lsim$ a few hundred km/s. The fraction of DM converted into relativistic energy is about $(1-exp(\Gamma^{-1}t(z))V_{k}/c$ $\approx 3\times 10^{-4}-10^{-5}$ with lifetime $\sim$ a few Gyr. Second, DDM results in significant free-streaming of daughter SDM particles. The free-streaming velocity of daughter particle suppresses structure growth on scales smaller than the free-streaming scale, an effect similar to that caused by massive standard model neutrinos or WDM. 
The temperatures of WDM and neutrinos decrease due to the expansion of the universe from the moment of decoupling, and so does their free-streaming effect. However, for DDM, the behavior of the free-streaming length can be very different, depending on the value of the decay lifetime. The Ly$\alpha$ forest systems probe the growth of structure at $z \sim 2-4$ and so in DDM models with decay lifetimes much smaller than a few Gyr (which is roughly the age of the universe at that time) all the DM particles will have decayed away by these redshifts. The free-streaming length will then start to shrink in a manner similar to WDM. On the other hand, if the decay lifetime is comparable to or larger than a few Gyr, the impact of DM peculiar velocities will become more significant at later times in the Ly$\alpha$ forest systems. A detailed exploration of the DDM free-streaming effects can be found in \cite{wang_zentner12}.

We have computed the linear matter power spectrum in DDM models using a modified {\tt CAMB} code. Examples of DDM power spectrum changes in the linear regime are shown in Figure~\ref{fig:PK_3d}, alongside spectrum changes derived from our N-body simulations. We plot the ratio of DDM to CDM power for two different choices of decay parameter combinations: (1) $V_{k}$ = 200 km/s, $\Gamma^{-1}$=10 Gyr; and (2) $V_{k}$ = 100 km/s, $\Gamma^{-1}$=1 Gyr. Changing $V_{k}$ or $\Gamma^{-1}$ has different effects on matter power suppression. We refer readers to \cite{wang_zentner12} for further discussion. We can see that the suppression is limited to small scales and that at large scales the DDM power remains the same as for CDM. This is necessary for the DDM model to solve the problems on galactic scales while at the same time agreeing with observations on large scales. Although the methods used to derive limits from different Ly$\alpha$ forest data differ, we can see that the analytical calculation broadly agrees with the numerical simulation results over the range of wavenumbers that the data probe. The light blue shaded area in Figure~\ref{fig:PK_3d} indicates the upper limits in wavenumber for the data sets that we consider. Although the Ly$\alpha$ forest already probes the quasi-linear region, at high redshift the growth of non-linear structure is mild. This explains the modest discrepancy shown in the comparison of simulation and analytical calculations in Figure~\ref{fig:PK_3d}. The matter power suppression scale is determined by the free-streaming length of the SDM particles. The decay lifetime changes the fraction of DM particles that possess the non-negligible peculiar velocities, and the kick velocity generated from the mass loss fraction in the decay process sets the scale of the DM peculiar velocities. We will see next in \S~\ref{subsection:flux power spectrum} that, based on the assumption that at high redshift baryons trace the DM distribution well, similar features  will also be found in the Ly$\alpha$ forest power spectrum.

\begin{figure*}[ht]
\includegraphics[height=12cm]{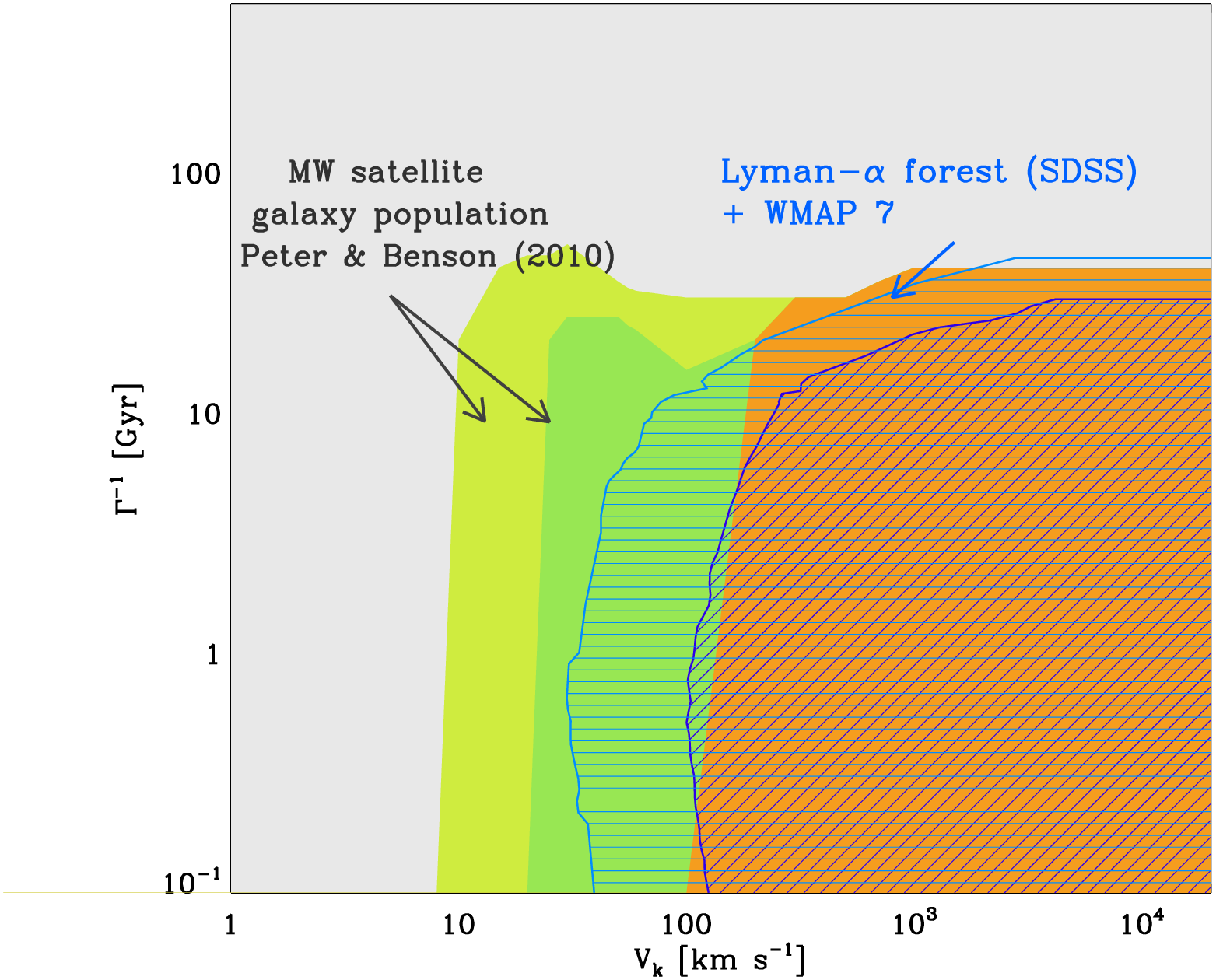}
\caption{ 
Comparison of DDM parameter 1 $\sigma$ exclusion contours from \citep{peter_etal10} (orange region on the right) and \citep{peter_benson10} (light and dark green on the left) to 
those that are derived from current Ly$\alpha$ forest data. The purple, diagonally-hatched region is the 1-$\sigma$ exclusion region from the 
VHS data set \cite{VHS_04} together with WMAP 7 data \citep{Komatsu_etal11} and the SDSS galaxy 3D power spectrum \citep{Tegmark_etal06}. 
The light blue line horizontally-hatched regions shows the 1-$\sigma$ exclusion contours from the SDSS 1D Ly$\alpha$ forest 
power spectrum \citep{McDonald_etal06}. The green regions depict those regions of parameters space that may strongly alter 
interpretations of the small-scale problems facing CDM and a significant portion of the green region is independently ruled out 
by our Ly$\alpha$ forest analysis (see text for details).
}
\label{fig:constraints}
\end{figure*}

\subsection{Signatures of Decaying Dark Matter in the 1D Lyman-$\alpha$ Forest Power Spectrum }
\label{subsection:flux power spectrum}

In Figure~\ref{fig:FP} we show the 1D Ly$\alpha$ forest  power spectra  as a function of redshift  derived from our numerical simulations. A caveat to these results is that, as we mentioned above, we have assumed that neutral hydrogen gas follows the underlying DM distribution. Although this is found to be a good approximation in CDM and WDM scenarios, it has not yet been extensively tested in the context of DDM models. However the similarities between DDM, WDM, and CWDM phenomenology suggest that this should be a useful approximation. We can imagine that this assumption will be most reliable in cases where decay lifetime is smaller than $\sim$~a~few~Gyr because decays take place before the formation of Ly$\alpha$ forest systems. So baryons will be more likely to be in gravitational equilibrium with DM, even if the latter have significant, non-thermal peculiar velocities due to the decay. 

Comparing Figure~\ref{fig:PK_3d} and Figure~\ref{fig:FP}, we can see that the behavior of the flux power spectrum due to decay is similar to what we observed in the matter power spectrum, as expected. In Figure~\ref{fig:FP} each set of points with the same color are an SDSS 1D flux power spectrum at a particular redshift bin. Along with each of the SDSS 1D flux power spectra are predictions derived from simulations (solid: CDM; dot-dashed: DDM). From top to bottom there are 11 uniformly spaced redshift bins from $z=4.2$ to $z=2.2$. The effective optical depth in each bin is fixed for both CDM and DDM simulations. This quantity affects the normalization of the flux power spectrum. In the left panel of Figure~\ref{fig:FP}, the effects of DDM on the flux power spectrum become more significant at lower redshift because decay processes become more common at lower redshifts. On the right the effect is already quite significant at redshift around $z \sim 4$ because the decay lifetime is comparable to the age of the universe at this redshift.

Figures~\ref{fig:PK_3d} and Figures~\ref{fig:FP} demonstrate that the Ly$\alpha$ forest is sensitive to the effects of DDM. We expect that experiments that probe structure growth at low redshift will be more effective in this class of decay models. In Ref.~\cite{wang_zentner12}, it was shown that using future or forthcoming weak lensing data we will be able to gain even better sensitivity than current Ly$\alpha$ forest data. However, the strength of these constraints will rely in part on theoretical modeling of the matter power spectrum on strongly non-linear scales and may be subject to significant uncertainty as a result. The advantage of the Ly$\alpha$ forest data (aside from being currently available) is that they probe comparably mild overdensities and are less challenging to treat theoretically.

\begin{figure*}[ht]
\includegraphics[height=12cm]{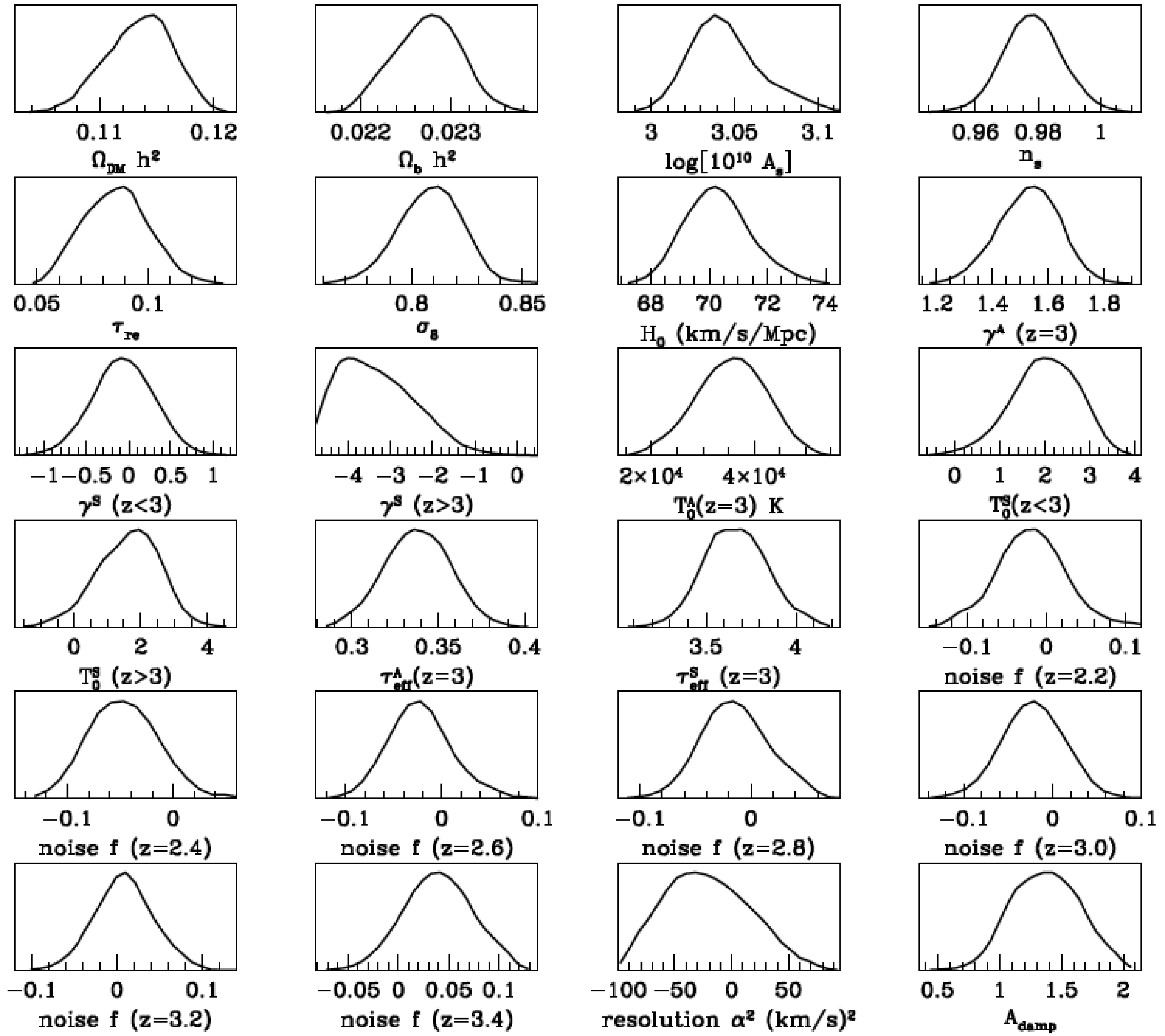}
\caption{ 
1D marginalized posterior probabilities for the cosmological and astrophysical parameters that we consider 
in our analysis of SDSS Ly$\alpha$ forest data. Cosmological and astrophysical parameters are inferred using 
a Taylor expansion of the flux power spectrum of the fiducial model to first order, based on our simulation suite, 
and performing MCMC analysis over the full parameter space.
}
\label{fig:MCMC}
\end{figure*}


\subsection{Lyman-$\alpha$ Forest Constraints on DDM Model Parameters}
\label{section:constraints}
In this section we present the results of our DDM limits using current Ly$\alpha$ forest data combined with WMAP 7 CMB anisotropy and 
SDSS 3D galaxy power spectrum data. Figure~\ref{fig:constraints} depicts the 1-$\sigma$ exclusion contours for DDM models 
using the VHS data set and the SDSS 1D flux power spectra. We display our $1\sigma$ exclusion 
contours alongside a variety of other contemporary constraints. The most relevant contemporary constraints come from 
modifications to the structures of DM halos with virial velocities similar to the SDM kick velocities \cite{peter_etal10} (orange region).  
Additional constraints may be placed on unstable dark matter by examining the properties of the satellite galaxies 
of the Milky Way \cite{peter_benson10} (light and dark green regions). However, these constraints rely on a variety of assumptions regarding the 
formation and evolution of relatively small galaxies. Alternatively, these constraints delineate a range of DDM parameters for which unstable 
dark matter may have a significant effect on the interpretation of the problems of the missing satellites of the Milky Way (e.g., \cite{klypin_etal99b,moore_etal99} 
and the densities of the brightest Milky Way satellite galaxies (e.g., \cite{Boylan-Kolchin_etal11}. As such, the green regions delineate that 
parameter range for which it is most interesting to develop independent constraints on unstable dark matter.

A significant advantage of the Ly$\alpha$ forest data is that they can extend 
constraints on the DDM kick velocity significantly, as is evident in Fig.~\ref{fig:constraints}. 
Moreover, these limits are not significantly affected by nonlinear structure growth modeling. 
In particular, they are not sensitive to uncertainties in the modeling of very small-scale structures, 
such as the properties of satellite galaxies round Milky Way-sized host galaxies. As a consequence, 
the Ly$\alpha$ forest constraints constitute independent, and robust lower limits to the constraining 
power of current large-scale surveys and begin to rule out parameter values of interest to the 
small-scale structure problems of CDM, which are the subject of so much contemporary research.

As indicated in Figure~\ref{fig:constraints}, the VHS data can already give interesting constraints on 
DDM that are competitive with contemporary bounds. For lifetimes $\Gamma^{-1} \lesssim 10$~Gyr, 
VHS data constrain the kick velocity to $V_k \lesssim 100~\mathrm{km/s}$. At highe kick velocity, 
the VHS data constrain the lifetime to exceed roughly $\Gamma^{-1} \gtrsim 30~\mathrm{Gyr}$. 
There is a slight shift in the slope of the Ly$\alpha$ forest constraint contours when the lifetime becomes of order 
$\Gamma^{-1} \sim \mathrm{a}\ \mathrm{few}\ \mathrm{Gyr}$. This turn-over reflects the turn-over in the free-streaming scale and was also observed in the lensing analysis of \cite{wang_zentner12}. In models with 
$\Gamma^{-1} \ll 1$~Gyr, the free-streaming scale is a decreasing function of time at the redshifts probed 
by the Ly$\alpha$ forest, whereas for larger lifetimes, the free-streaming scale is increasing with time due 
to continuous decays at the epoch of the observation.

The SDSS data yield considerably stronger constraints than the VHS data (see Fig.~\ref{fig:constraints}) and, indeed, 
cut significantly into the region in which DDM models may alleviate some of the small-scale issues of CDM. Roughly 
speaking, our SDSS Ly$\alpha$ forest constraints extend the constraint on the kick velocity (or equivalently, the dark matter 
particle mass splitting) by more than half an order of magnitude to $V_k \lesssim 30~\mathrm{km/s}$ for lifetimes $\Gamma^{-1} \ll H_{0}^{-1}$ significantly. For higher kick velocities, the constraint is effective up to particle lifetime of $\Gamma^{-1} \sim 40~\mathrm{Gyr}$. 
 As with the VHS constraint, the constraining 
power dies quickly for longer lifetimes. The constraining power of these Ly$\alpha$ forest observations 
is fundamentally limited by the fact that the forest is observed at high-redshift and therefore cannot probe 
exceptionally long lifetimes very effectively.

In \cite{wang_zentner12} it was shown that future weak lensing surveys could achieve similar 
sensitivity to these Ly$\alpha$ forest constraints when limited to scales on which a linear perturbative theoretical analysis 
suffices. In particular, projected constraints on DDM kick velocities are quite similar. However, lensing data can improve 
constraints primarily through their ability to probe models with significantly larger lifetimes, $\Gamma^{-1} \gtrsim 30 H_0^{-1}$. 
This is a result of the distant galaxies being observed in contemporary and forthcoming imaging surveys being lensed 
significantly by structure at lower redshifts than are probed through the Ly$\alpha$ forest.

There are a number of systematic uncertainties which might affect our results. For example, 
as we mentioned earlier, the Ly$\alpha$ forest power spectrum bias function has not been tested and calibrated 
extensively within the DDM scenario. However,
based on the findings of \cite{Viel_etal05} with WDM models, 
we expect that it is likely to make only a negligible difference on our 
parameter constraints. Ref.~\cite{Viel_etal05} found that, given the large systematic plus 
statistical error bars on the VHS data, tests showed the suppression
of power in the WDM scenario has a negligible effect on 
on the bias function in the relevant k-space range. 
In our VHS analysis, we assume that the same behavior also holds for the DDM class of models based 
on the similarity between WDM and DDM power suppression. We leave further tests of the bias function 
using hydrodynamical simulations for future work, if necessary.

Other possible concerns arise from the fact that we adopt the {\it fluctuating Gunn-Peterson approximation} to derive flux power spectrum in dark-matter-only simulations instead of using full hydrodynamical simulations. In Figure~\ref{fig:MCMC} we show the 1D marginalized likelihood functions for our cosmological and astrophysical parameters derived from joint SDSS flux power spectrum and WMAP 7 data using MCMC methods. For standard cosmological parameters, the sensitivities are driven by CMB data and the results agree very well with published WMAP 7 values \citep{Komatsu_etal11}. Furthermore, we compare our astrophysics and nuisance parameter limits to those derived using hydrodynamic simulations by \cite{Viel_etal06} as shown in their Fig. 4. {\em Without priors}, our limits are quite 
similar to their results. For $\gamma$, the normalization and evolution parameters agree remarkably well. The width of our posterior likelihood function is likely narrower because of the inclusion of WMAP 7 data. For $\tau_{eff}$, the best-fit values also agree well but with a slope that is slightly greater in our case. For $T_0$, the discrepancy is larger. While the amplitude, $T_0^A$ (z=3), and the slope at z $<$ 3, $T_0^S$(z$<$3), agree well with those in \cite{Viel_etal06}, the slope at $ z>3$, $T_0^S(z>3)$, has different sign. From observations, the evolution of the temperature at the mean density $T_0$ may have a broken power law behavior \cite{Schaye_etal00}. However, the SDSS data error bars at high redshift are large, which gives loose constraints on IGM properties. Also it has been pointed out in \cite{Viel_etal06} that simulations usually generate higher temperature predictions than observational data. Therefore this discrepancy should have negligible effects on our results. 

\section{Discussion and Conclusion}
\label{section:conclusion}

In this paper we have explored the power of current Ly$\alpha$ forest data 
to place constraints on DDM lifetimes and mass splittings. The mass difference 
is parameterized by the velocity kick ($V_k$) that the daughter SDM particles 
receive upon the decay of the heavier parent DDM particles. The kick speed is 
related to the mass splitting in the non-relativistic limit by $V_k = c\, \Delta M/M$. 
DDM leads to a suppression of matter clustering on scales below the free-streaming 
scale of the daughter SDM particles and this suppression can be probed with observed 
Ly$\alpha$ forest power spectrum data.

We have considered two Ly$\alpha$ forest data sets: the VHS \cite{VHS_04} and SDSS data \cite{McDonald_etal06}. The strongest 
limits come from the SDSS 1D flux power spectrum because of the very small statistical error bars on the range of wavenumbers that 
we make use of. Most of the constraining power comes from data at low redshift bins. Although the original 1D flux power spectrum in the VHS 
data probes smaller scales than the SDSS data set, we take a more conservative upper wavenumber limit ($k \lsim 0.02724~\mathrm{s/km}$) 
for the derived matter power spectrum in our MCMC analysis. It may be possible to construct more restrictive constraints using the 
flux power spectra from \cite{kim_etal04,Croft_etal02} directly as was attempted in Ref.~\cite{Seljak_etal06}; however, we leave such an effort 
for future work using high-resolution hydrodynamic simulations to calibrate the IGM properties at small scales.

We find that the VHS data exclude kick velocities $v_k \gtrsim 100-230$~km/s for $\Gamma^{-1} \lsim 10$~Gyr, 
a result that is competitive with contemporary constraints \cite{peter_etal10}. The SDSS data, which 
we combine with WMAP 7 data, place limits of $v_k \gtrsim 30-70$~km/s for $\Gamma^{-1}\lsim 10$~Gyr. 
High kick velocities (large mass splittings) can only be accommodated if the lifetime of the DDM particle is $\Gamma^{-1} \gtrsim 30 -40~\mathrm{Gyr}$. These new bounds significantly 
extend existing model-independent bounds on dark matter decaying to invisible species. 

These constraints are interesting because they restrict parameters for which 
the effects of DDM on the Milky Way satellite galaxy population should be
important.  It may be possible to achieve similar constraints depending upon a variety of assumptions 
regarding the formation process of these satellite galaxies \cite{peter_benson10}, but
the Ly$\alpha$ forest provides a complementary constraint using data at different redshifts, on different 
length scales, in the mildly non-linear regime (as opposed to the extreme non-linear regime of 
dwarf galaxy formation). The Ly$\alpha$ forest provides another independent, competitive constraint
which is most competitive for small decay lifetimes ($\Gamma^{-1} \lsim H_0^{-1}$) because it probes structure 
growth at early time. Where the Ly$\alpha$ forest constraints are least restrictive, at large lifetimes, forthcoming weak lensing survey 
may have the ability to improve upon contemporary constraints. We point readers to \cite{wang_zentner12} for further discussion 
of the advantages and limitations of studying DDM models using weak lensing methods. 

In closing, we have demonstrated that measurements of the large-scale matter distribution
made with Ly$\alpha$ forest surveys can be a powerful probe of DDM. Our constraints 
are already among the most stringent constraints on unstable dark matter that do not rely on any assumptions 
regarding the decay products and eliminate a significant portion of the interesting parameter space for 
DDM models. There are a number of reasons to believe that similar methods will yield 
even more restrictive constraints in the coming year. 
First, the precision of the BOSS 1D Ly$\alpha$ forest power spectrum data \cite{Palanque-Delabrouille_etal13} will be greatly improved compared 
to that from earlier datasets due to their large sample. Second, \cite{Bird_etal11b} demonstrated that BOSS data may be sufficiently 
powerful on their own to break degeneracies between the IGM and cosmological parameters. Furthermore, 
the recently released WMAP Nine-Year data \cite{WMAP9} and PLANCK data \cite{Planck_13} provide higher precision 
cosmological parameter measurements using CMB anisotropy and will more strongly constrain the cosmological parameters. 
A joint analysis of the new CMB data with BOSS Ly$\alpha$ forest data may greatly improve our understanding of matter density 
fluctuations at high redshift, which will provide better limits on unstable DM properties, among other novel physics. Fully 
exploiting these high-quality data will likely require a significant simulation effort to ensure the robustness of any constraints 
derived from BOSS data. In addition to the constraints that we present in this manuscript, it is our hope that our work will 
motivate more detailed numerical studies of novel dark matter models with the goal of improving constraints from 
Ly$\alpha$ absorption data as well as additional possible constraints from related observations. 



%
%
\vspace*{12pt}
\begin{acknowledgments}
We are grateful to Tiziana Di Matteo, Jeff Newman, and Louis Strigari for useful comments and discussions. We also thank Bryan Arant for helping modifying the Gadget-2 code. MW and ARZ were funded by the the US National Science Foundation through grant PHY 0968888 and by the Pittsburgh Particle Physics, Astrophysics, and Cosmology Center (PITT PACC) at the University of Pittsburgh. The work of ARZ was also supported in part by the National Science Foundation under Grant No. PHYS-1066293 and the hospitality of the Aspen Center for Physics. RACC acknowledges support from NSF award AST 1109730. AHGP was supported by a Gary McCue Fellowship through the Center for Cosmology at UC Irvine, NASA Grant No. NNX09AD09G and NSF grant 0855462. Simulations were performed in the Frank supercomputer of the Center for Simulation $\&$ Modeling (SaM) at the University of Pittsburgh.
\end{acknowledgments}


\bibliography{lyaddm}


\end{document}